\documentclass[aip,nofootinbib]{revtex4}

\usepackage{epsfig}
\usepackage[fleqn,sumlimits,intlimits]{amsmath}
\usepackage{latexsym,amssymb}
\usepackage[english]{babel}
\usepackage{verbatim}

\def\half{\frac{1}{2}}

\def\on#1#2{{\buildrel{\mkern2.5mu#1\mkern-2.5mu}\over{#2}}}
\newcommand{\beq}{\begin{equation}}
\newcommand{\eeq}{\end{equation}}

\begin{document}

\title{Localization of particles in quantum field theory}

\author{Willem Westra}
\affiliation{Division of Mathematics, The Science Institute, University of Iceland,
Dunhaga 3, 107 Reykjavik, Iceland \\
\texttt{\textup{w.westra@raunvis.hi.is}}}

\begin{abstract}
We put forward an interpretation of scalar quantum field theory as relativistic quantum mechanics by curing well known problems related to locality.  A probabilistic interpretation of quantum field theory similar to quantum mechanics is difficult if particle localization is defined using the Newton-Wigner position operator as it is  non-local and non-covariant. An alternative bilinear covariant position operator is discussed which incorporates a time operator that can be exponentiated to a unitary operator. Moreover, it satisfies an algebra that unifies special relativity and quantum mechanics and has the same form for particles with spin. Higher power position operators are derived which yield Heisenberg's uncertainty relations. Our ideas are illustrated with a relativistic wave function whose probability density can be perfectly localized.
\end{abstract}

\maketitle

%%%%%%%%%%%%%%%%%%%%%%%%%%%%%%%%%%%%%%%%%%%%%%%%%%%%%%%%%%%%%%%%%%%%
\section{Outline}
%%%%%%%%%%%%%%%%%%%%%%%%%%%%%%%%%%%%%%%%%%%%%%%%%%%%%%%%%%%%%%%%%%%%
%
In this paper we present a consistent first quantization of scalar quantum field theory. The term ``first quantization'' refers to the formulation of relativistic quantum field theory as  relativistic quantum mechanics with a Fock space of single particle wave functions that have a Copenhagen-like probabilistic interpretation. In appendix \ref{Conventional}, we briefly address the standard relativistic quantum mechanics interpretation of quantum field theory and its 
problems. We recall that in standard textbook treatments on scattering theory single particle wave functions are identified with positive frequency solutions of the Klein Gordon equation. The restriction to positive frequency solutions implies that the time evolution is governed by the Feynman propagator. Even though the wave function interpretation of positive frequency solutions is used to compute scattering probabilities and cross sections this interpretation is generally not deemed fundamental as it suffers from two well known problems related to non-locality,
\begin{enumerate}
\item Time evolution as described by the Feynman propagator violates Einstein causality.  \\
In other words, it causes wave functions to spread outside of the lightcone. 
\item The conventional Newton-Wigner position operator \cite{Newton:1949cq} is inherently non-local and non-covariant. 
\end{enumerate}
 In appendix \ref{Conventional} we review the first problem and in appendix \ref{causality} we summarize recent work \cite{causalalternative} which resolves it. We show that adding negative frequency modes to the wave function restores Einstein causality, precisely as in classical field theory. As in \cite{causalalternative} we 
emphasize that negative frequency modes actually have positive energy as dictated by the energy momentum tensor. The time ordered time evolution of wave functions which include negative frequency modes is governed by the half-advanced half-retarded Wheeler propagator and therefore replaces the Feynman propagator\footnote{Within our conventions the Wheeler propagator is the real part of the Feynman propagator and it is also known as the ``principal part propagator'', see for instance \cite{Greiner:1996zu}, due the appearance of the principal part in its Fourier representation \cite{Greiner:1996zu, Birrell:1982ix, Itzykson:1980rh}. The Wheeler propagator can also be thought of as being half the time ordered commutator \cite{causalalternative}.}. The inclusion of negative energy modes moreover implies that the Klein Gordon field itself is interpreted as a wave function. We do so by recalling a form of the probability density from \cite{causalalternative} that was inspired by \cite{Halliwell:1992nj} and \cite{Henneaux:1982ma} which is constructed from the field and the Hadamard two point function\footnote{In the cosmological literature the Hadamard function is referred to as the statistical propagator, and in second quantization it is given by the anti-commutator. As observed in \cite{Halliwell:1992nj}, we emphasize however that the Hadamard function is not a propagator in the sense that it does not govern the dynamical propagation of a field in time, see sections \ref{negfreqprob} and \ref{Hilbert} of this paper.}. 

Our emphasis on causal methods for quantum field theory is partially motivated by the pivotal role causality plays in a recent nonperturbative approach to quantum gravity called Causal Dynamical Triangulations (CDT). This method was introduced in \cite{Ambjorn:1998xu} where two dimensional world sheets were quantized while preserving a causal foliation. This work has been generalized to $3+1$ dimensions using computer simulations, for recent work see for instance \cite{Ambjorn:2010hu,Ambjorn:2010fv}. Additionally, the causal world sheet perspective was further developed to include interactions by topology changes \cite{Ambjorn:2009wi,Ambjorn:2009fm}.

In section \ref{Hilbert} we introduce an, as far as we are know, new and very simple form of the probability density which is constructed from the field and its Hilbert transform, a transform that is widely used in signal analysis. We interpret this new incarnation of the probability density as a representation in a different basis of the positive/negative frequency form implied by the innerproduct presented by Woodard \cite{Woodard:1989ac} and Halliwell-Ortiz \cite{Halliwell:1992nj}.

In section \ref{localization} we address the second source of non-locality in standard quantum field theory which stems from the localization of relativistic particles through the Newton-Wigner position operator \cite{Newton:1949cq}. We discuss an alternative bilinear position operator which shares many of the desirable properties of the Newton-Wigner operator but does not possess any non-local terms, both in momentum and position space. Furthermore, the bilinear position operator is part of a covariant space-time position operator, unlike the Newton-Wigner operator.  Additionally the bilinear operator is not by itself related to dynamics and is therefore a generalization of the Schr\"{o}dinger picture position operator in non-relativistic quantum mechanics. The Newton-Wigner operator on the other hand satisfies a Heisenberg like equation. 

A particularly distinguishing feature of the bilinear operator is that it generalizes to higher powers of the coordinates. Such higher power position operators which are Hermitian with respect to relativistically normalized wave functions have to the best of our knowledge not been derived before. They allow us to re-derive the Heisenberg uncertainty relations from quantum field theory viewed as relativistic quantum mechanics. It is quite remarkable that this corner stone of quantum mechanics is no longer present in quantum field theory if one insists on using the Newton-Wigner operator. Consequently, many authors have abandoned a quantum mechanics interpretation of quantum field theory entirely, see for instance \cite{Halvorson:2001hb, Peres:2002wx, Haag:1992hx, Hartle:1992as,Marolf:2002ve}. We show that using the bilinear position operators on wave functions that include negative frequencies restores the quantum mechanical interpretation of free quantum field theory. It is expected that a similar picture to holds in interacting field theory provided the interaction can be treated perturbatively.

To conclude our discussion on the position operator, the observation is made that unlike the Newton-Wigner operator the bilinear operator is purely a position operator in the sense that it does not contain any other generators of the Lorentz group. This has the important consequence that the form of the operator does not depend on the spin or mass of the fields under consideration.

In section \ref{Localized states} we discuss a real solution to the Klein Gordon equation that is normalized with respect to our probability density using the Hilbert transform. It behaves qualitatively similar to a gaussian wave function and can therefore be perfectly localized if the width approaches zero, where the width is defined using our bilinear position and position squared operators. Infinitely precise localization requires infinite energy however, which is consistent with physical expectations related to Heisenberg's uncertainty relations. We contrast our states with the ``localized states'' of Newton and Wigner \cite{Newton:1949cq} which are neither normalizable nor perfectly localizable. 
%

%%%%%%%%%%%%%%%%%%%%%%%%%%%%%%%%%%%%%%%%%%%%%%%%
\section{The Hilbert transform and the probability density} \label{Hilbert}
%%%%%%%%%%%%%%%%%%%%%%%%%%%%%%%%%%%%%%%%%%%%%%%%%%%%%%%%%%%%%%%%%%%%
The position space form of the probability density that is used in textbook treatments of scattering probabilities can be written as follows\footnote{For a discussion of the probabilistic interpretation used in conventional scattering theory see appendix \ref{Conventional}.},
\beq \label{KGprobability}
P(x) = \phi_+^*(x) i \on{\leftrightarrow}{\partial}_{0} \phi_+(x).
\eeq
We refer to this as the Klein Gordon probability density. If one adds negative frequency modes\footnote{For details on the addition of negative frequency modes see appendix \ref{causality}.} following \cite{Henneaux:1982ma} and \cite{Halliwell:1992nj} one can find the equivalent form, 
\beq \label{plusminprobability}
P(x) =
\half\left( \phi_+^*(x) i \on{\leftrightarrow}{\partial}_{0} \phi_+(x) 
- \phi_-^*(x)  i \on{\leftrightarrow}{\partial}_{0}  \phi_-(x)\right).
\eeq
As discussed in \cite{causalalternative}, the Hadamard function can be used rewrite this in a manifestly real probability density on the space of real solutions of the Klein Gordon equation in position space. 
\beq \label{Hadamardprobabilitydensity}
P(x) = 
\half\int d^3x' 
\phi(x) \on{\leftrightarrow}{\partial }_{t} \Delta_H(x,x') \on{\leftrightarrow}{\partial }_{t'} \phi(x')\Big|_{t=t'},
\eeq
The Hadamard form of the probability density \eqref{Hadamardprobabilitydensity} contains an integral however which makes it slightly unnatural as a Born-like probability. 
%This integral can be performed in the positive/negative frequency decomposition to obtain \eqref
%{plusminprobability}. However, the frequency decomposition obscures the connection with real valued fields and their dynamics as described by the Wheeler %propagator. 
Therefore we present an, as far as we know, new form of the probability density. Similar to the Hadamard form, it is defined in terms of real fields but does not contain an integral in its definition\footnote{A similar form of the innerproduct is discussed in the context of PT-symmetric quantum mechanics by A. Mostafazadeh \cite{Mostafazadeh:2006fz}. Instead of Hilbert's integral transform the author utilizes a non-local differential operator $\hat{C}$ which is only 
defined for free theories. The operator $\hat{C}$ can be related \cite{Kleefeld:2006bp} to the charge grading operator that appears in the inner product of PT-symmetric quantum mechanics introduced by Bender, Brody and Jones \cite{Bender:2002vv}. Because of the similarities between the Hilbert transform and $\hat{C}$ one might therefore deduce that also the Hilbert transform can be given such an interpretation.},
\beq \label{Hilbertprobability}
P(x) = \half\phi(x) \on{\leftrightarrow}{\partial}_0 \phi_H(x),
\eeq
where $\phi_H(x)$ denotes the Hilbert transform of $\phi$ in the time variable. 
The Hilbert transform can be written as a principal part integral 
%,
\beq
 \phi_H = \frac{1}{\pi} P \int^{\infty}_{-\infty} dt'\frac{\phi(t',x)}{t-t'},
\eeq
or more explicitly as
\beq \label{realhilbert}
\phi_H = \lim_{\epsilon \rightarrow 0} \frac{1}{\pi} \int^\infty_\epsilon dt'\frac{\phi(t+t',x) - \phi(t-t',x)}{t'}. 
\eeq
Note that with these definitions the Hilbert transform is performed in the time direction. One can nonetheless extend the Hilbert transform in a Lorentz covariant manner to any time like direction, see \cite{Kaiser:1990ww}. Lorentz invariance of the Hilbert transform can furthermore be inferred from the fact that it only affects the sign of the energy modes of the field, which inherently is a Lorentz invariant operation.

The position space representation of the probability density \eqref{Hilbertprobability} for real fields has a simple, manifestly real form. At first sight it seems surprising that we can define a probability density entirely with real variables. A closer look however, reveals that although the definition of the Hilbert transform \eqref{realhilbert} does not involve complex numbers it defines a complex structure on the space of solutions in the sense that if it is applied twice it gives back the negative of the original function,
\beq
\hat{J}_H^2 \phi = - \phi, \qquad \hat{J}_H \phi= \phi_H. 
\eeq
The Hilbert transform is well known in signal analysis and one of it's essential qualities is that it flips the sign of the negative frequency modes of a signal. 
\beq
\phi_H = i(\phi_+ -\phi_-) = -\Im \phi_+,\qquad\qquad \phi =\phi_+ + \phi_- = \Re \phi_+.
\eeq
From this property one can easily see that our definition of the innerproduct \label{hilbertinnerproduct} is equivalent to the form \eqref{plusminprobability}. 
The Fourier decomposition of the  field and its Hilbert transformed field can thus be written as,
\beq
\phi(x) =\int \frac{\mathrm{d}^4k}{(2 \pi)^3}\delta(k^2+m^2)f(k)e^{i k x} \qquad
\phi_H(x) = i\int \frac{\mathrm{d}^4k}{(2 \pi)^3}\delta(k^2+m^2)\epsilon(k^0)f(k)e^{i k x},  
\eeq
where $f(k)$ is the wave packet in momentum space and if we define the following
\beq
\phi_H(x) =
 \int \frac{\mathrm{d}^4k}{(2 \pi)^3}\delta(k^2+m^2)f_H(k)e^{i k x} =
 \int \frac{\mathrm{d}^4k}{(2 \pi)^3}\delta(k^2+m^2)f^*_H(k)e^{-i k x},
\eeq
we see that
\beq
f_H(k) = i\epsilon(k^0)f(k),\qquad \qquad f^*_H(k) = -i\epsilon(k^0)f^*(k).
\eeq
Our conventions for the Hilbert transform are such that $\hat{J}_H \cos(\omega t) = \sin(\omega t)$. We stress that the Hilbert transform of a real signal is again a real signal, which means that $\phi_H$ is real. This should be clear from the fact that it can be written as minus the imaginary part of the positive frequency field. The rationale is often actually reversed and the Hilbert transform is used to, uniquely, define the complex valued positive frequency signal from the real valued input.  From this point of view the positive frequency component  $\phi_+(x)$ can be considered as an analytic signal representation of the field $\phi$.  

From the Hadamard form of the probability density \eqref{Hadamardprobabilitydensity}  it is clear that the Hilbert transform of the field can be obtained by acting on it with the Hadamard propagator,
\beq
\phi_H(x) = \int d^3x' \Delta_H(x,x')\on{\leftrightarrow}{\partial }_{t'}\phi(x') ,
\eeq
We gave here a presentation of our probability density \eqref{Hilbertprobability} as a manifestly real form of \eqref{plusminprobability} which is the negative frequency generalization of the Klein Gordon density \eqref{KGprobability}. One does not need to make such an interpretation related to Fourier modes however, one can just take our probability formula \eqref{Hilbertprobability} at face value. It is possible to compute the time evolution of the field through the Wheeler propagator \eqref{Wheelerpropagator}, compute the Hilbert transform of the time evolved field and insert both in our probability density \eqref{Hilbertprobability}. In this way \emph{one can formulate quantum mechanics entirely without complex valued wave functions\footnote{The wave functions are real for neutral scalar fields, charged scalar fields are modeled as complex fields.}}. We stress furthermore that from the same point of view the Klein Gordon equation for real scalar fields \emph{is} the relativistic generalization of the Schr\"{o}dinger equation. Moreover, since the time evolution is governed by the Wheeler propagator instead of the Feynman propagator the evolution is manifestly causal, see \cite{causalalternative} and appendix \ref{causality}.

%%%%%%%%%%%%%%%%%%%%%%%%%%%%%%%%%%%%%%%%%%%%%%%%%%%%%%%%%%%%%%%%%%%%
\subsection{Particles and anti particles: complex scalar fields}
%%%%%%%%%%%%%%%%%%%%%%%%%%%%%%%%%%%%%%%%%%%%%%%%%%%%%%%%%%%%%%%%%%%%
In the discussion above we focused on the real scalar field and how it can be interpreted as a real valued quantum mechanical wave function. To properly discuss particles and antiparticles however it is instructive to examine a complex scalar field. We do so by writing the complex field as a combination of two real fields, 
\beq
\varphi = \phi_1+i \phi_2.
\eeq
Within our formalism $\varphi$ and $\varphi^*$ are the wave functions for a charged particle and an anti-particle respectively. The main physical difference with the real scalar field is that while a real field is neutral, a complex field carries electric charge and the charge density is given by
\beq \label{chargedensityrealfields}
\rho =
\varphi^*(x) i \on{\leftrightarrow}{\partial}_t \varphi(x) 
= \phi_2 \on{\leftrightarrow}{\partial}_t \phi_1
- \phi_1 \on{\leftrightarrow}{\partial}_t \phi_2 
= 2  \phi_2 \on{\leftrightarrow}{\partial}_t \phi_1,
\eeq
where the diagonal terms have disappeared because the $\phi_1$ and $\phi_2$ fields are electrically neutral separately. The naive generalization of our probability density of real fields \eqref{Hilbertprobability} is not real,
\beq \label{naiveprobability}
\varphi^*(x) \on{\leftrightarrow}{\partial}_t \varphi_H(x)  
=  \half\left(\phi_1 \on{\leftrightarrow}{\partial}_t \phi^H_1
+
\phi_2 \on{\leftrightarrow}{\partial}_t \phi^H_2 
+ 
i\phi_1 \on{\leftrightarrow}{\partial}_t \phi^H_2-i\phi_2 \on{\leftrightarrow}{\partial}_t \phi^H_1\right).
\eeq
The proper probability density for complex scalar fields is just the real part of the above, 
\beq \label{probabilitycomplex}
P(x) =\Re\varphi^*(x) \on{\leftrightarrow}{\partial}_t \varphi_H(x) =  
\half
\left(
\phi_1 \on{\leftrightarrow}{\partial}_t \phi^H_1  
+
\phi_2 \on{\leftrightarrow}{\partial}_t \phi^H_2
\right),
\eeq
in this form the probability density is manifestly real as required for a probability density. If we alternatively write it terms of positive and negative frequency wave functions we see that it has the exact same form as for real fields,
\beq
P(x) = \half\left(\varphi_+^*(x) i \on{\leftrightarrow}{\partial}_{t} \varphi_+(x) - \varphi_-^*(x)  i \on{\leftrightarrow}{\partial}_{t}  \varphi_-(x)\right),
\eeq
and is the generalization of the Klein Gordon probability density for complex fields,
\beq
P_{KG}(x) = \varphi_+^*(x) i \on{\leftrightarrow}{\partial}_{t} \varphi_+(x). 
\eeq
A consistency check that our probability density formula \eqref{probabilitycomplex} is indeed a probability density for a charged particle is that when replacing $\varphi$ with its charge conjugate $\varphi^*$ the density is left invariant. In other words, the probability is the same for positively and negatively charged particles as it should be. Similarly, acting with the same charge conjugation operation on our formula for the charge density for wavefunctions that include negative energy modes \eqref{chargedensityrealfields} has the physically required effect of changing its sign. 

In this subsection we have constructed a manifestly real probability density on the space of complex solutions to the Klein Gordon equation that includes negative energies. Particularly, in our formulation \emph{both particles and anti-particles are modeled on a Hilbert space that includes negative frequencies}. We therefore take the point of view that our construction is an alternative to the Feynman Stueckelberg interpretation of conventional ``in-out'' quantum field theory. Our approach is similar to standard quantum field theory in the sense that anti-particles are implemented as complex conjugates of particles but we do not need to interpret the antiparticles as positive frequency modes that ``travel back in time''.  Both the particle field/wave function $\varphi(x)$ and the antiparticle field/wave function $\varphi^*(x)$ are propagated with the Wheeler propagator exactly as in the case of real scalar fields. In other words, the time evolution of both particles and antiparticles is computed with the same half-advanced half-retarded boundary conditions.

%%%%%%%%%%%%%%%%%%%%%%%%%%%%%%%%%%%%%%%%%%%%%%%%%%%%%%%%%%%%%%%%%%%%
\section{Localization} \label{localization}
%%%%%%%%%%%%%%%%%%%%%%%%%%%%%%%%%%%%%%%%%%%%%%%%%%%%%%%%%%%%%%%%%%%%
Locality is generally considered to be a pivotal aspect of quantum field theory. The Feynman propagator is not a local object however. As stressed by for instance Hegerfeldt  and Ruijsenaars \cite{Hegerfeldt:1998ar,Ruijsenaars:1981sm, Hegerfeldt:1974qu}, the time evolution of a positive frequency wave function as encoded in the Feynman propagator is such that if the state is well localized initially it fills the whole of space an infinitesimal time later. This is one of the issues that has led to quite radical views that field theory cannot be considered as the relativistic quantum mechanics of localizable particles, see for instance  \cite{Halvorson:2001hb, Peres:2002wx, Haag:1992hx, Hartle:1992as}. As explained above however we consider the whole real field to be the wave function. The corresponding time evolution is described by the real Klein Gordon equation and the Wheeler propagator and is therefore perfectly causal which implies that no infinitely fast spreading of wave functions can occur, see particularly \cite{causalalternative}. Hence we deduce that introducing negative frequency solutions and writing the probability density in terms of the real field itself \eqref{Hilbertprobability} cures the non-locality problem associated with the a-causal spreading of wave functions. However, we remind the reader that the non-locality of the Feynman propagator is not a problem of standard quantum field theory itself  but is purely a problem with its ``in-out'' description, the ``in-in''  formalism \cite{Calzetta:1986ey, Jordan:1986ug, Chou:1984es, Weinberg:2005vy, Berges:2004yj} is causal \cite{causalalternative,Park:2010pj,Soussa:2003vn, Koksma:2007uq} although not always manifestly so, see for instance the Feynman diagram like formulations \cite{Musso:2006pt, vanderMeulen:2007ah, Giddings:2010ui}. For applications see e.g.  \cite{Prokopec:2002uw, Prokopec:2002jn, Prokopec:2003qd, Koksma:2009wa}.

%%%%%%%%%%%%%%%%%%%%%%%%%%%%%%%%%%%%%%%%%%%%%%%%%%%%%%%%%%%%%%%%%%%%
\subsection{The Newton-Wigner position operator}
%%%%%%%%%%%%%%%%%%%%%%%%%%%%%%%%%%%%%%%%%%%%%%%%%%%%%%%%%%%%%%%%%%%%  
  
While viewing the real field itself as the wave function eliminates the non-local spreading of wave functions, another source of non-locality remains that obstructs a proper first quantized picture. This other non-locality is hidden in the localization problem associated with the relativistic quantum mechanics interpretation of 
quantum field theory. The problem is to find a suitable position operator that can be used to compute the average position given a specific quantum state. In a famous paper Newton and Wigner \cite{Newton:1949cq} introduced a position operator that is named after the authors\footnote{Pryce was actually the first \cite{Pryce:1948pf} to introduce the Newton-Wigner position operator in the context of photons, but since Newton and Wigner gave a systematic derivation from seemingly reasonable axioms their names became associated with the operator.}. The Newton-Wigner operator is defined on the space of positive energy solutions and therefore inherently involves infinitely fast spreading wave packets. It is possible to define analogous position operators on the Hilbert space that includes negative energy solutions, see for example \cite{Mostafazadeh:2006fz}. Such operators still possess non-local behavior that is inherent to the Newton-Wigner operator though.

Before discussing our position operator we start by a derivation of the Newton-Wigner position operator for scalar particles which differs considerably from the presentation by Newton and Wigner but is very concise and logically consistent. We list the well known problems associated with the Newton-Wigner operator.  Then we derive a position operator that satisfies some of the desirable properties of the Newton-Wigner operator but does not share its problems and furthermore provides a covariant extension.

If one takes the positive frequency Klein Gordon density \eqref{KGprobability} seriously as a probability density one is inclined to define the position operator through the following expectation value,
\beq \label{xvecexpec}
\langle \vec{x} \rangle=
\half\int d^3x~ \vec{x} ~\phi^*_+(x) i \on{\leftrightarrow}{\partial}_{0} \phi_+(x) ,
\eeq
this definition seems obvious but poses problems when going to momentum space. The naive position operator that realizes the above expectation value in momentum space is precisely the non-relativistic operator,
\beq
\hat{\vec{x}}_{NR} \rightarrow i \partial_{\vec{k}},
\eeq
so that the expectation value in momentum space is given by,
\beq \label{nr}
\langle \hat{\vec{x}}\rangle =  \int \frac{\mathrm{d}^3k}{(2 \pi)^3 2 \omega_k }
 \Big[\phi^*_+(k)  i \partial_{\vec{k}}\,\phi_+(k)\Big]. 
\eeq
Because of the non-trivial measure factor $(2 \omega_k)^{-1}$ in \eqref{nr} the non-relativistic position operator is obviously not Hermitian.  The Newton-Wigner operator is simply the Hermitian part of the non-relativistic operator,
\beq \label{NewtonWigner}
\hat{\vec{x}}_{NW} \rightarrow
\frac{i}{2} (\partial_{\vec{k}} + \partial_{\vec{k}}^\dag ) 
= i (\partial_{\vec{k}} +\half \partial_{\vec{k}}\frac{1}{\omega_k})= 
i (\partial_{\vec{k}} - \frac{\vec{k}}{2\omega^2_k}).
\eeq
Hence this operator contains a non-trivial function of the momentum, which is at the heart  of the problems of the Newton-Wigner position operator. First of all this term prevents a simple interpretation of the position operator in position space, its representation in position space does \emph{not} lead to the simple expectation value \eqref{xvecexpec}. Instead the operator becomes highly non-local in position space\footnote{Formula \eqref{NewtonWignerposition} is the differential analogue of the integral formula presented by Newton and Wigner \cite{Newton:1949cq}.},
\beq \label{NewtonWignerposition}
\hat{\vec{x}}_{NW} \rightarrow (\vec{x}+\half\frac{\partial_{\vec{x}}}{\partial_{\vec{x}}^2+m^2}).
\eeq
As is widely discussed in the literature this non-local behavior makes the interpretation of the Newton-Wigner operator as an exact position operator highly questionable. The expression is moreover not covariant and transforms non-trivially under Lorentz transformations, so to what extent it can be interpreted as a relativistic operator is not clear. We will not discuss particles with spin in depth in this article but we just mention that the Newton-Wigner operator does not even exist for massless particles with spin greater then one-half such as photons. Generalizations of the Newton-Wigner operator exist that apply to photons but are not physically acceptable since they either have non-commuting components or do not transform as three vectors. 

The absence of acceptable relativistic position operators presents us with an awkward situation. Planck and Einstein realized that electromagnetic radiation should be emitted in finite energy wave-packets to resolve the ultra violet catastrophe of the Rayleigh-Jeans law.  The reason \emph{why} the electromagnetic field should be subdivided into quanta is explained by the statistical interpretation of Einstein and Born. The fact that the square of the field amplitudes should be interpreted as a probability density enforces the fields to ``square integrate'' to multiples of one, which is a constraint that supplements the ``classical'' field equations. The unavailability of suitable relativistic position operators however implies that there is no probabilistic interpretation for relativistic fields that is similar to quantum mechanics. Often this fact is considered to be a mere curiosity but it should be realized that the absence of a first quantized Copenhagen like interpretation of relativistic fields removes the naturalness of the partitioning of Klein Gordon fields, Dirac fields and Maxwell fields into packets as required by the photoelectric effect and the Bose-Einstein, Fermi-Dirac, and Planck distributions respectively. 

In the next section \emph{we introduce a  position operator that is covariant, local, causal and furthermore commutes with the spin generators and is therefore also valid for fields with spin}. Such an operator restores the probability interpretation of non-relativistic quantum mechanics in a relativistic setting and therefore also restores the naturalness of the subdivision of fields into quanta.  
%
%%%%%%%%%%%%%%%%%%%%%%%%%%%%%%%%%%%%%%%%%%%%%%%%%%%%%%%%%%%%%%%%%%%%
\subsection{The bilinear space-time operator}
%%%%%%%%%%%%%%%%%%%%%%%%%%%%%%%%%%%%%%%%%%%%%%%%%%%%%%%%%%%%%%%%%%%% 
We simply define the space-time position operator from the expectation value of its non operator valued components in position space,  
\beq \label{xoperatorposition}
\langle \hat{x}^\mu \rangle=
\half\int d^3x~ x^\mu ~\phi(x)  \on{\leftrightarrow}{\partial}_{0} \phi_H(x). 
\eeq
Note that we implement our probability density which includes negative energy modes to maintain causality, the computation is quite similar however if one chooses a positive frequency Hilbert space.  We use the following off shell Fourier transforms, 
\beq \label{Fouriertransform1}
\phi(x) = \int \frac{\mathrm{d}^4k}{(2 \pi)^4}\phi(k)e^{i k x} \,, \qquad \phi_H(x)
 = \int \frac{\mathrm{d}^4k'}{(2 \pi)^4}\phi^*_H(k')e^{-i k' x}.
\eeq
where we have that
\beq
\phi^*_H(k) = -i2\pi\delta(k^2+m^2)\epsilon(k^0)f^*(k).
\eeq
Let us start with the derivation of the position operator. The essential step that avoids the problems of the Newton-Wigner operator is to represent the $x^\mu$ components in \eqref{xoperatorposition} by a bilinear derivative in momentum space,
\beq \label{bilinearxposition}
\langle \hat{\vec{x}} \rangle=\frac{1}{4}\int d^3x\frac{\mathrm{d}^4k}{(2 \pi)^4}\frac{\mathrm{d}^4k'}{(2 \pi)^4} 
\phi(k)\phi^*_H(k')
\left(k_0+k'_0\right) 
e^{i(k_0 -k_0')t}
(i\partial_{\vec{k}'} - i\partial_{\vec{k}})e^{i( \vec{k} -\vec{k}')\cdot\vec{x}}.
\eeq
Upon evaluation of the on shell delta functions and using partial integration the expectation value can be written as follows,
\beq
\langle \hat{\vec{x}} \rangle =   \frac{1}{4}\int \frac{\mathrm{d}^3k}{(2 \pi)^3 2 \omega_k} 
 \left[
  f^*(k_+,t)i\on{\leftrightarrow}{\partial}_{\vec{k}} f(k_+,t)+
   f^*(k_-,t)i\on{\leftrightarrow}{\partial}_{\vec{k}} f(k_-,t) 
  \right],
\eeq
which can be written in a covariant manner as,
\beq \label{topexp}
\langle \hat{\vec{x}} \rangle= \frac{i}{2}\int \frac{\mathrm{d}^4k}{(2 \pi)^3} \delta(k^2+m^2)
\left[f^*(k,t)\on{\leftrightarrow}{\partial}_{\vec{k}} f(k,t)\right],
\qquad
f(k,t) = f(k) e^{ik_0 t}.
\eeq
Here we stress that in the above expression the integral over the frequency $k^0$ should be performed \emph{before} taking the derivative. Hence, the position operator is given by,
\beq \label{positiononlyoperator}
\hat{\vec{x}}\rightarrow  \frac{i}{2}\on{\leftrightarrow}{\partial}_{\vec{k}},
\eeq
For general wave functions we can see that the average position, or ``center of probability'', is given by,
\beq \label{trajectory}
\langle \hat{\vec{x}} \rangle(t) = \langle \hat{\vec{x}} \rangle(0) +  \langle \hat{\vec{v}}_g \rangle \: t.
\eeq
Observe that the motion of the wave packet is a solution to Newton's law of motion for particles where an acceleration term is absent since the wave packet is free. The initial location of the ``center of probability'' of the wave packet is given by,
\beq
\langle \hat{\vec{x}} \rangle =   \frac{1}{4}\int \frac{\mathrm{d}^3k}{(2 \pi)^3 2 \omega_k} 
 \left[
  f^*(k_+)i\on{\leftrightarrow}{\partial}_{\vec{k}} f(k_+)+
   f^*(k_-)i\on{\leftrightarrow}{\partial}_{\vec{k}} f(k_-) 
  \right],
\eeq
and the velocity of the wavepacket is the expectation value of the group velocity,
\beq \label{groupvelocity}
\langle \hat{\vec{v}}_g \rangle = \half\int \frac{\mathrm{d}^3k}{(2 \pi)^3 2 \omega_k }
 \frac{\partial \omega_k}{\partial \vec{k}}( |f(k_+)|^2 - |f(k_-)|^2).
\eeq
Interestingly, formula \eqref{groupvelocity} allows us to understand the group velocity as an operator in the momentum space representation of wave functions. Equation \eqref{trajectory} shows that the ``center of probability'' of general wave packets follow a particle like trajectory which should be interpreted in a similar spirit as Ehrenfest's theorem.

Because we simply define a covariant space-time operator from the expectation value \eqref{xoperatorposition} we also need to promote time to a quantum mechanical operator. While conflicting with popular belief, time can be promoted to an operator quite easily, provided one takes a Lorentz covariant perspective\footnote{As operators time and position are on equal footing but the expectation values of these operators are certainly \emph{not} on the same footing. Accordingly, the energy time Heisenberg relation $\Delta E \Delta t \gtrsim \half\hbar$ does not have the same exact status as the momentum position uncertainty relation $\Delta P \Delta x \geq  \half\hbar$. The asymmetry can be traced to the fact that the integration in expectation values is over a \emph{spacelike} hypersurface in a nevertheless covariant manner.}. If one chooses to Fourier transform all four coordinates to compute the expectation value, time will automatically need to be represented as an operator.
\beq \label{bilinearxposition}
\langle \hat{\tilde{t}} \rangle=\frac{1}{4}\int d^3x\frac{\mathrm{d}^4k}{(2 \pi)^4}\frac{\mathrm{d}^4k'}{(2 \pi)^4} 
\phi(k)\phi^*_H(k')
\left(k_0+k'_0\right) 
e^{i( \vec{k} -\vec{k}')\cdot\vec{x}}
(i\partial_{k_0'} - i\partial_{k_0})e^{i(k_0 -k_0')t}
\eeq
If one evaluates the above expression one simply obtains, 
\beq
\langle \hat{\tilde{t}} \rangle = t, 
\eeq
where $t$ is simply the time coordinate, or equivalently the time parameter, that labels the spatial hypersurfaces. The result is simple because we chose our spacelike hypersurfaces to be surfaces of constant coordinate time $t$. One is however free to use different hypersurfaces, then the expectation value of the time operator would be non-trivial and give the average coordinate time of the hypersurfaces. Note that this use of the time operator is not directly related to time of arrival problems\footnote{Our time operator can also be used to study time of arrival type problems if one uses a different recipe for computing expectation values, this is however beyond the scope of the current paper.}. By partial integration we see that the expectation value of the time operator can be written as
\beq
\langle \hat{\tilde{t}} \rangle = -\frac{1}{4}\int \frac{\mathrm{d}^4k}{(2 \pi)^3} \delta(k^2+m^2)
\left[ f^*(k,t)i\on{\leftrightarrow}{\partial}_{k_0} f(k,t)-f^*(k)i\on{\leftrightarrow}{\partial}_{k_0} f(k) \right], 
\eeq
where we stress that unlike in the case of the position operator one should evaluate the on shell delta functions \emph{after} taking the derivative. We read the above formula as, $\langle \hat{\tilde{t}} \rangle_t = \langle \hat{t} \rangle_t  -\langle \hat{t} \rangle_{t=0}$ such that the time operator is given by
\beq \label{timeoperator}
\hat{t}\rightarrow  -\frac{i}{2}\on{\leftrightarrow}{\partial}_{k_0},
\eeq
and $t_0 = \langle \hat{t} \rangle_{t=0} $ is the value of time at the spatial hypersurface for which $t=0$, where the time operator is defined such that
\beq \label{topexp}
\langle \hat{t} \rangle_t= t + t_0= -\frac{1}{4}\int \frac{\mathrm{d}^4k}{(2 \pi)^3} \delta(k^2+m^2)
\left[f^*(k,t)i\on{\leftrightarrow}{\partial}_{k_0} f(k,t)\right]
\eeq
Curiously there is a relative sign if one compares the time operator with the position operator. Combining the above results \eqref{positiononlyoperator} and \eqref{timeoperator}  defines the covariant bilinear space-time position operator,
\beq \label{truepositionoperator}
\hat{x}^\mu \rightarrow  \frac{i}{2}\on{\leftrightarrow}{\eth}_{k_\mu}, 
\qquad 
\on{\leftrightarrow}{\eth}_{k_0} \rightarrow -\on{\leftrightarrow}{\partial}_{k_0},
\qquad
\on{\leftrightarrow}{\eth}_{\vec{k}} \rightarrow \on{\leftrightarrow}{\partial}_{\vec{k}}
\eeq
and the expectation value of the space-time position operator is given by,
\beq
\langle \hat{x}^\mu \rangle = \frac{1}{4}\int \frac{\mathrm{d}^4k}{(2 \pi)^3} \delta(k^2+m^2)
\left[f^*(k,t)i\on{\leftrightarrow}{\eth}_{k_\mu} f(k,t)\right]
\eeq
Where again we emphasize that, when one computes the position expectation value, one should first perform the frequency integral and then apply the derivative while for the time operator the order should be reversed. Furthermore  it is important that when we write the bilinear derivatives we understand them to act on the wave packets only and not on the delta function. Interestingly, we have represented time both by an operator and by a parameter, the two properties are not mutually exclusive as is sometimes argued. The expectation value of time as an operator simply returns the value of the time parameter that labels the spatial hypersurfaces. The bilinear action avoids the problematic non-local and non covariant terms in the Newton-Wigner operator  (\ref{NewtonWigner}, \ref{NewtonWignerposition}). 

A similar bilinear position operator was, as far as we know, introduced into the conventional positive frequency formalism of quantum field theory by Olkhovsky and Recami, see the for example the recent paper \cite{Recami:2010zz} and the references therein. Our operator differs however from that of the aforementioned authors in two essential aspects.  Firstly, they interpret their bilinear operator as the antisymmetric part of a total non covariant Newton-Wigner like operator. The antisymmetric position operator is considered to be a ``mean position operator'' while the symmetric operator describes in a rather opaque manner the width of a relativistic wavepacket. One problem with that point of view is that it leads to a peculiar modification of Heisenberg's uncertainty relation. We on the other hand show that the width and other characteristics of wave packets can be computed completely according to the standard rules of quantum mechanics by the standard deviation and higher order moments. In particular, we show below that our interpretation leads to the same form of the Heisenberg uncertainty relations as in non-relativistic quantum mechanics. 

The second aspect where our operator differs is that our time operator is defined as the conjugate of the off shell frequency and not as the conjugate of the (on shell) energy. One implication of this is that even for systems with discrete energy levels time will be represented by a continuous operator. Another consequence is the different sign of the time operator and position operators. Moreover, since our time operator is dual to the frequency and we include negative frequency modes in our wave functions the exponentiated time operator is a unitary translation operator. Time operators that are dual to the energy cannot simply be exponentiated to a unitary operator since the energy is bounded from below. This unitarity problem plagues most other proposals for time operators and was discussed initially by Pauli \cite{Pauliunitarity}.  In other words, not only causality but also unitarity indicates that wave functions that include negative frequency modes should be used .

%%%%%%%%%%%%%%%%%%%%%%%%%%%%%%%%%%%%%%%%%%%%%%%%%%%%%%%%%%%%%%%%%%%%
\subsection{Powers of the space-time operator}
%%%%%%%%%%%%%%%%%%%%%%%%%%%%%%%%%%%%%%%%%%%%%%%%%%%%%%%%%%%%%%%%%%%% 

We define the ``position-squared operator'' from the expectation value,
\beq
\langle \hat{x}^i \hat{x}^j \rangle=\half\int d^3x~ x^i x^j ~\phi(x)  \on{\leftrightarrow}{\partial}_{0} \phi_H(x), 
\eeq
where the indices $i,j$ are the spatial components of $x^\mu$. 
The position components can similar to before be given a bilinear representation in momentum space,
\beq 
\label{kkkk}
\langle \hat{x}^i \hat{x}^j \rangle= 
-\frac{1}{4}\int \frac{\mathrm{d}^4k}{(2 \pi)^3} \delta(k^2+m^2)
f^*(k,t)
\left(
\on{\leftarrow}{\eth}_{k_i}\on{\leftarrow}{\eth}_{k_j}
+
\on{\rightarrow}{\eth}_{k_i} \on{\rightarrow}{\eth}_{k_j}
\right)
f(k,t),
\eeq
where it should be noted that one should evaluate the $k_0$ integral before taking the derivative. Equation \eqref{kkkk} yields an, as far as we know new, real and symmetric bilinear position squared operator,
\beq
\widehat{x^i x^j}= 
-\half\left(\on{\leftarrow}{\eth}_{k_i}\on{\leftarrow}{\eth}_{k_j}
+\on{\rightarrow}{\eth}_{k_i} \on{\rightarrow}{\eth}_{k_j}\right).
\eeq
Using $f(k,t) = f(k)e^{ik_0t}$ we see that the dynamics of the position squared operator at non zero times behaves as follows,
\beq
\langle \hat{x}^i\hat{x}^j \rangle(t) = 
\langle \hat{x}^i \hat{x}^j \rangle(0) 
+ ( \langle \hat{v}^i_g \hat{x}^j \rangle +\langle \hat{v}^j_g \hat{x}^i \rangle) \: t 
+  \langle \hat{v}^i_g \hat{v}^j_g \rangle \: t^2,
\eeq
where the initial expectation value is given by,
\beq
\langle \hat{x}^i \hat{x}^j \rangle(0) = 
-\frac{1}{4}\int \frac{\mathrm{d}^3k}{(2 \pi)^3 2\omega_k} 
\left[
f^*(k_+)\left(\on{\leftarrow}{\eth}_{\! \vec{k}_i}\on{\leftarrow}{\eth}_{\! \vec{k}_j}
+
\on{\rightarrow}{\eth}_{\! \vec{k}_i}\on{\rightarrow}{\eth}_{\! \vec{k}_j} \right)f(k_+) 
\right]+  (+ \rightarrow -),
\eeq
The coefficients of the terms linear in time are of the form,
\beq
\langle \hat{v}^i_g \hat{x}^j \rangle=
\frac{1}{4}\int \frac{\mathrm{d}^3k}{(2 \pi)^3 2 \omega_k}
\left[
\frac{\partial \omega_{k}}{\partial k_i }f^*_+(k)i \on{\leftrightarrow}{\eth}_{k_j}f_+(k)
\right]
-  (+ \rightarrow -),
\eeq
and the term quadratic in time is the expectation value of the group velocity operator squared,
\beq
\langle \hat{v}^i_g \hat{v}_g^j \rangle=
\frac{1}{2}\int \frac{\mathrm{d}^3k}{(2 \pi)^3 2 \omega_k}
\left[
\frac{\partial \omega_{k}}{\partial k_i } \frac{\partial \omega_{k}}{\partial k_j } f^*_+(k)f_+(k)
\right]
+  (+ \rightarrow -),
\eeq
With the position operator and the position squared operator at hand we can compute the dynamics of the width of a general wave packet. The width in the $z$ direction is for example given by,
\beq
\Delta z^2(t) = \langle \hat{z}^2 \rangle -\langle \hat{z} \rangle^2  +  
2(\langle \hat{v}_g^z \hat{z} \rangle -  \langle \hat{v}^z_g \rangle\langle z \rangle)~t + 
( \langle (\hat{v}^z_g)^2 \rangle -\langle \hat{v}^z_g \rangle^2)~t^2,
\eeq
where the averages are defined with respect to the wave packet at $t=0$ which is $f(k)$.

We define the time squared operator in a similar fashion as the position squared operator,
\beq 
\label{kk}
\langle \widehat{t^2} \rangle= 
-\frac{1}{4}\int \frac{\mathrm{d}^4k}{(2 \pi)^3} \delta(k^2+m^2)
f^*(k,t)\left((\on{\leftarrow}{\eth}_{k_0})^2+(\on{\rightarrow}{\eth}_{k_0})^2\right)f(k,t),
\eeq
this can be seen to give,
\beq
\langle \widehat{t^2} \rangle = (t+t_0)^2,
\eeq
where we have that the initial time parameter $t_0$ is consistent with our time operator,
\beq
t_0 = -\frac{1}{4}\int \frac{\mathrm{d}^4k}{(2 \pi)^3} \delta(k^2+m^2) 
f^*(k)i\on{\leftrightarrow}{\partial}_{k_0} f(k),
\eeq
and we have that, in addition the fact that $t^2_0$ is simply the square of the above,
\beq \label{t0sq}
t^2_0 =  -\frac{1}{4}\int \frac{\mathrm{d}^4k}{(2 \pi)^3} \delta(k^2+m^2) 
f^*(k)\left((\on{\leftarrow}{\eth}_{k_0})^2+(\on{\rightarrow}{\eth}_{k_0}\right)^2 f(k).
\eeq
This follows from our requirement that in position space,
\beq
\langle \widehat{t^2} \rangle  
=\half\int d^3x~ (t+t_0)^2 ~\phi(x)  \on{\leftrightarrow}{\partial}_{0} \phi_H(x) = (t+t_0)^2,
\eeq
which for $t=0$ directly implies \eqref{t0sq}. 

The time squared and position squared operators can be combined in a covariant operator
\beq
\widehat{x^\mu x^\nu}= 
-\half\left(\on{\leftarrow}{\eth}_{\! \vec{k}_\mu}\on{\leftarrow}{\eth}_{\! \vec{k}_\nu}
+
\on{\rightarrow}{\eth}_{\! \vec{k}_\mu}\on{\rightarrow}{\eth}_{\! \vec{k}_\nu} \right).
\eeq
The covariant position operator and its higher moments possess many desirable properties. One important aspect is that they make sense as quantum operators since they are Hermitian. The Hermiticity is even manifest as the position operator clearly is imaginary and antisymmetric and the position squared operator is 
real and symmetric. 

%%%%%%%%%%%%%%%%%%%%%%%%%%%%%%%%%%%%%%%%%%%%%%%%%%%%%%%%%%%%%%%%%%%%
\subsection{The Weyl Heisenberg Poincar\'{e} algebra}
%%%%%%%%%%%%%%%%%%%%%%%%%%%%%%%%%%%%%%%%%%%%%%%%%%%%%%%%%%%%%%%%%%%% 
The bilinear position operator is part of a remarkable algebra that embodies both quantum mechanics and special relativity without resorting to ``second quantization''. The position operator satisfies covariant commutation relations,
\beq \label{comm1}
[\hat{x}^\mu, \hat{x}^\nu]=  [\hat{p}_\mu, \hat{p}_\nu]= 0, 
\eeq
\beq  \label{comm2}
[\hat{x}^\mu, \hat{p}_\nu]= i\delta^\mu_\nu \hat{I}, 
\eeq
This algebra is a relativistic generalization of the Weyl-Heisenberg algebra. To derive the commutation relations test functions on the left as well as to the right of the operators are needed since in our setting operators are understood to have a bilinear action. The identity generator is related to phase transformations and is a central element of the algebra since it commutes with all other generators,
\beq  \label{comm3}
[\hat{x}^\mu,  \hat{I}]= [\hat{p}_\mu,  \hat{I}]= [\hat{L}_{\mu\nu}, \hat{I}]=0. 
\eeq
The bilinear position operator also gives a differential operator representation of the Poincar\'{e} algebra. Translations in momentum space are encoded in the relations,
\beq  \label{comm4}
[\hat{L}_{\mu\nu}, \hat{x}_\rho]=  \eta_{\mu\rho}\, \hat{x}_\nu -\eta_{\nu\rho} \hat{x}_\mu, 
\eeq
and the standard Poincar\'{e} algebra is given by, 
\beq  \label{comm5}
[\hat{L}_{\mu\nu}, \hat{p}_\rho]=  \eta_{\mu\rho}\, \hat{p}_\nu -\eta_{\nu\rho} \hat{p}_\mu, 
\eeq
\beq \label{Lorentzalgebra}
[\hat{L}_{\mu_1\nu_1},\hat{L}_{\mu_2\nu_2}]=  
\eta_{\mu_1\mu_2}\,\hat{L}_{\nu_1\nu_2}
 -\eta_{\mu_1\nu_2}\,\hat{L}_{\nu_1\mu_2}
 - \eta_{\nu_1\mu_2}\,\hat{L}_{\mu_1\nu_2}+ \eta_{\nu_1\nu_2}\,\hat{L}_{\mu_1\mu_2}. 
\eeq
The complete algebra might be named the ``Weyl-Heisenberg-Poincar\'{e}'' algebra. It is the algebra of ``events'' in space-time and is the commutative version of Snyder's algebra \cite{Snyder:1947}, which is well known in the field of non-commutative geometry, see for instance \cite{Elias:2006ec, Battisti:2010sr}. In this sense our position operator unifies the coordinate concept of special relativity with the position concept of relativistic quantum mechanics. If one uses the Newton-Wigner position operator on the other hand one needs to clearly distinguish those two concepts as recognized for example in \cite{Toller:1998wf}. An important lesson that we extract from the bilinear position operator is that the differential operator representation of the Weyl-Heisenberg-Poincar\'{e} algebra 
should be taken seriously as a defining characteristic of relativistic quantum mechanics since it can be consistently realized by Hermitian operators both in momentum as in position space. Moreover, by exponentiation of the generators, the Poincar\'e group can be unitarily represented in momentum and position space\footnote{As discussed above, even our time operator can be exponentiated to give a unitary operator since it is dual to the frequency, not the energy. Since the frequency unlike the energy takes on values over the whole real line, it is not bounded from below, therefore our time operator does not suffer from the unitarity problem addressed by Pauli \cite{Pauliunitarity}. }.
The standard realization of the generators in position space is
\beq \label{generatorsposition}
\hat{L}_{\mu\nu} =  
\hat{x}_\mu \hat{p}_\nu -\hat{p}_\nu \hat{x}_\mu, \qquad \hat{x}_\mu 
\rightarrow x_\mu, \qquad \hat{p}_\mu \rightarrow -i\partial_\mu. 
\eeq
The differential representation of the generators in position space \eqref{generatorsposition} is well known and their algebra is sometimes interpreted in a 
quantum mechanical sense but one encounters problems in momentum space if one uses the Newton-Wigner position operator. It does not satisfy the covariant commutation relations the algebra cannot be realized in momentum space. Hence, a complete relativistic quantum interpretation of the Weyl-Heisenberg-Poincar\'{e} algebra cannot be given if one adheres to Newton-Wigner localization. The bilinear position operator \emph{does} allow us to represent the algebra in momentum space however,
\beq
\hat{L}_{\mu\nu} =  \hat{x}_\mu \hat{p}_\nu -\hat{p}_\nu x_\mu, \qquad \hat{x}_\mu \rightarrow \frac{i}{2}\on{\leftrightarrow}{\eth}_{k_\mu}, \qquad \hat{p}_\mu \rightarrow p_\mu
\eeq
Consequently, this realization completes the single particle relativistic quantum mechanical interpretation. We emphasize again that the bilinear position operator should act on wave functions that include negative frequencies, otherwise problems with non-locality will remain if one considers amplitudes, and the time operator could not be exponentiated to a unitary operator. We also point out that observables do not depend on whether one discusses these in position or in momentum space.

%%%%%%%%%%%%%%%%%%%%%%%%%%%%%%%%%%%%%%%%%%%%%%%%%%%%%%%%%%%%%%%%%%%%
\subsubsection*{Particles with spin}
%%%%%%%%%%%%%%%%%%%%%%%%%%%%%%%%%%%%%%%%%%%%%%%%%%%%%%%%%%%%%%%%%%%%

At the level of the algebra it is not difficult to introduce spin by adding to the generators of orbital Lorentz transformations the spin generators of local Lorentz transformations\footnote{It is sometimes asserted that spin is an inherently quantum phenomenon. We do not agree with this statement and simply view spin as the conserved charge related to active Lorentz transformations in the rest frame, i.e. the local Lorentz transformations.}. 
\beq
\hat{J}_{\mu\nu}=\hat{L}_{\mu\nu}+\hat{S}_{\mu\nu}.
\eeq
If one replaces the orbital Lorentz generators $\hat{L}$ in (\ref{comm1} -\ref{Lorentzalgebra}) with the total angular momentum operator $\hat{J}_{\mu\nu}$ one obtains the correct algebra for spinning relativistic particles. The spin generators seperately generate a Lorentz algebra which is formally equivalent to \eqref{Lorentzalgebra}, and commute with the orbital generators, 
\beq
[\hat{L}_{\mu\nu},\hat{S}_{\rho\sigma}]=0.
\eeq
The local nature of the spin generators is stressed by the fact that they commute with translations and coordinates,
\beq
[\hat{S}_{\mu\nu},\hat{p}_\rho]=0, \qquad [\hat{S}_{\mu\nu},\hat{x}^\rho]=0.
\eeq
These commutation relations also make it clear that \emph{the bilinear position operator is defined independent of the spin and mass of a quantum field}. This is diametrically opposite to for example the Newton-Wigner operator for the Dirac field and the Pryce operator for the Maxwell field \cite{Pryce:1948pf}
 which depend non-trivially on the spin generators. Pryce's operator for the photon is for example closely related to the spinless Newton-Wigner operator \eqref
{NewtonWignerposition},
\beq
\hat{x} \rightarrow i (\partial_{\vec{k}} - \frac{\vec{k}}{2\omega^2_k})  + \frac{\vec{k} \times \hat{s}}{\omega^2_k},
\eeq
where in this case $\omega_k = \sqrt{\vec{k}^2}$ since photons are massless. The components of the local Lorentz algebra valued vector $\hat{s}$ are the spin one  generators of the $so(3)$ sub algebra given by the epsilon symbol $(s_i)_{j k} = \varepsilon_{ijk}$. Correspondingly, the Pryce position operator interacts non-trivially with the spin of the photon. Moreover, the non-trivial spin dependence causes the unphysical effect that the Pryce coordinates of a photon do not commute, similar non-commutativity occurs for other operators as well, see for instance \cite{JaekelReynaud1, Jaekel:1997hs}.

As emphasized above, the bilinear position operator does not act non-trivially on the local Lorentz spin indices, in fact it does not contain any of the other 
generators of the Lorentz group. It therefore carries no dynamical information which implies that \emph{the bilinear position operator can be consistently 
understood as an operator in the Schr\"{o}dinger picture while the Newton-Wigner operator behaves like an operator in the Heisenberg picture.} This is 
exemplified by the fact that the commutation relation of the Newton-Wigner position operator with the zero component of the momentum has the form of a 
Heisenberg equation,
\beq
\frac{\hat{p}_i}{\hat{p}^0} =\hat{v}_i= i [\hat{p}_0,\hat{x}^{NW}_i] ,
\eeq
where $\hat{v}$ denotes the Newton-Wigner velocity operator. The bilinear three-position operator is time independent on the other hand and commutes with the zero component of the momentum as demanded by the Weyl-Heisenberg-Poincar\'{e} algebra. Covariance of the bilinear position operator formalism can therefore be understood as coming from a clear separation between kinematics and dynamics.  The kinematics is covariantly described by the Weyl-Heisenberg-Poincar\'{e} algebra and the dynamics is described in a covariant manner by the Klein Gordon, Dirac and Maxwell equations of motion.     

%%%%%%%%%%%%%%%%%%%%%%%%%%%%%%%%%%%%%%%%%%%%%%%%%%%%%%%%%%%%%%%%%%%%
\subsection{The Heisenberg uncertainty relation in relativistic quantum mechanics}
%%%%%%%%%%%%%%%%%%%%%%%%%%%%%%%%%%%%%%%%%%%%%%%%%%%%%%%%%%%%%%%%%%%%
Since we have constructed a position squared operator as well as a position operator we are able to compute the coordinate space width and the momentum width of a wave function as described by the respective standard deviations. Using these standard deviations we can derive Heisenberg's uncertainty relations for relativistic quantum mechanics in almost the same manner as in non-relativistic quantum mechanics. To make the presentation as clear as possible we present Heisenberg's uncertainty relation for covariant relativistic wave functions in $1+1$ dimensions. The average position and the average position squared 
are simply given by
\beq
\langle x \rangle=\half\int dx~ x ~\phi(x)  \on{\leftrightarrow}{\partial}_{0} \phi_H(x), \qquad 
\langle x^2 \rangle=\half\int dx~ x^2 ~\phi(x)  \on{\leftrightarrow}{\partial}_{0} \phi_H(x),
\eeq
Upon inserting the Fourier representation of the field we obtain
\beq 
\langle x \rangle = 
\frac{i}{2}\int \frac{d p}{(2 \pi)2\omega_k} f^*_+(p) \on{\leftrightarrow}{\eth}_{p}  f_+(p) +~(+\leftrightarrow -), \qquad
\langle x^2 \rangle = -\frac{1}{2}\int \frac{d p}{(2 \pi)2\omega_k} f^*_+(p)\big(\on{\leftrightarrow}{\eth}_{k}\big)^2  f_+(p) +~(+\leftrightarrow -),
\eeq
and the analogous width of the momentum space wave packet is defined by the momentum expectation values,
\beq 
\langle k \rangle = i\int \frac{d p}{(2 \pi)2\omega_k}~k~\left(|f_+(k)|^2 + |f_-(k)|^2\right), \qquad
\langle k^2 \rangle = i\int \frac{d p}{(2 \pi)2\omega_k} ~k^2~\left(|f_+(k)|^2 + |f_-(k)|^2\right),
\eeq
the standard deviations are defined as usual,
\beq
(\Delta x)^2 = 
\langle x^2 \rangle - \langle x \rangle^2 \qquad (\Delta p)^2 
= \langle k^2 \rangle - \langle k \rangle^2 .
\eeq
The first step to obtain Heisenberg's inequality one uses the Cauchy Schwarz inequality. The inequality holds exactly as in standard non-relativistic quantum mechanics, the non-trivial relativistic measure does not lead to alterations, 
\beq
\langle x^2 \rangle \langle k^2\rangle \geq \langle x k \rangle^2.
\eeq
Also the second step goes through exactly as in non-relativistic quantum mechanics. The expectation value of $\hat{x}\hat{p}$ is always bigger than the 
expectation value of its imaginary part $ -i\half[\hat{x},\hat{k}]$, 
\beq
\langle x k \rangle^2 \geq  \frac{1}{4}\langle \| [\hat{x},\hat{k}] \| \rangle^2,
\eeq
shifting the operators on the left by the average produces $\Delta x^2 \Delta k^2$ and leaves the commutator invariant. Again exactly as for the position operator of non-relativistic quantum mechanics we have the commutation relation $[\hat{x},\hat{k}]=i$ and taking the square root on both sides we have that Heisenberg's uncertainty principle holds precisely as in non-relativistic quantum mechanics,
\beq
\Delta x \Delta k  \geq  \frac{1}{2}.
\eeq
%

%%%%%%%%%%%%%%%%%%%%%%%%%%%%%%%%%%%%%%%%%%%%%%%%%%%%%%%%%%%%%%%%%%%%
\subsubsection*{Equivalence to Schr\"{o}dinger position operators}
%%%%%%%%%%%%%%%%%%%%%%%%%%%%%%%%%%%%%%%%%%%%%%%%%%%%%%%%%%%%%%%%%%%%
In momentum space one can relate our wave functions $f(k)$, which are Lorentz scalars, to wave functions that are normalized as in non relativistic quantum mechanics,
\beq
\psi_+(\vec{k}) = 
\frac{1}{\sqrt{(2 \pi) 2\omega_k}} f(+\omega_k,\vec{k}), \qquad \psi_-(\vec{k}) 
= \frac{1}{\sqrt{(2 \pi) 2\omega_k}} f(-\omega_k,\vec{k}).
\eeq
With such a redefinition it is not immediately obvious that our position operators  are equivalent with the position operators of non relativistic quantum mechanics. A short calculation shows however that
\beq \label{relnonrel}
\frac{i}{2}\frac{1}{(2 \pi)2\omega_k} 
f^*(\pm \omega_k, \vec{k}) \on{\leftrightarrow}{\eth}_{k} f(\pm\omega_k,\vec{k})
=
 \frac{i}{2}\left(\psi_\pm^*(\vec{k}) \partial_{k} \psi_\pm(\vec{k})\right),
\eeq
which follows by antisymmetry of the bilinear derivative, partial integration and the definition of the $\eth$ operator. The right hand side of \eqref{relnonrel} has the form appropriate for a standard quantum mechanical expectation value. The position squared operator is also equivalent to the standard operator of non relativistic quantum mechanics. Up to certain boundary terms that vanish at infinity we have that,
\beq
\frac{i}{2}\frac{1}{(2 \pi)2\omega_k} 
f^*(\pm \omega_k, \vec{k})
\left(\on{\leftarrow}{\eth^2}_{k}
+ \on{\rightarrow}{\eth^2}_{k}\right) 
f(\pm\omega_k,\vec{k})
=
\psi_\pm^*(\vec{k})\partial^2_{k} \psi_\pm(\vec{k}).
\eeq
So we conclude that the expectation values of our relativistic position operators can be written in the same form as in non-relativistic quantum mechanics.  Hence it is not surprising that Heisenberg's uncertainty relation is exactly the same in the relativistic and non-relativistic cases. Note that we kept the negative frequency contributions explicit however. 
\section{Localized states} \label{Localized states}
%%%%%%%%%%%%%%%%%%%%%%%%%%%%%%%%%%%%%%%%%%%%%%%%%%%%%%%%%%%%%%%%%%%%
An important part of the work by Newton and Wigner \cite{Newton:1949cq} is their definition of a so called ``localized'' state. Their ``localized'' state appears as an eigenstate of the Newton-Wigner position operator. Notice that we used quotation marks to indicate that the Newton-Wigner wave functions are in fact only approximately localized. In momentum space the Newton-Wigner ``localized'' wave function is given by,
\beq
f^+_{NW}(\omega_k,\vec{k})= \sqrt{\omega_k} 
= \sqrt[4]{\vec{k}^2+m^2}.
\eeq
Besides being not perfectly localized the Newton-Wigner wave function is also not a proper wave function since it is not normalizable. This was already noted by Newton and Wigner and is obvious if one inserts in the Klein Gordon normalization formula \eqref{KGnormalization}. The position space form of the wave function makes it clear that it is not perfectly localized,
\beq
f^+_{NW}(t=0,\vec{x})
= \frac{1}{2^\frac{7}{4} \pi^\frac{3}{2} \Gamma(\frac{1}{4})} \left(\frac{m}{r}\right)^{5/4}K_{5/4}(mr),
 \qquad r=\|\vec{x}\|, 
\eeq
around $r =0$ the Newton-Wigner state behaves as $\frac{1}{r^{5/2}}$. This behavior remains valid in the massless limit,
\beq
f^+_{NW}(t=0,\vec{x})= \frac{1}{2^\frac{7}{2} \pi^{\frac{3}{2}}} \left(\frac{1}{r}\right)^{5/2},
\eeq
Note furthermore that the Newton-Wigner wave function is a positive frequency solution to the Klein Gordon equation.  

In \cite{causalalternative} we introduced a wave function for massless scalar particles that unlike the Newton-Wigner wave function is real, normalizable and perfectly localized in the sense that in a certain limit the probability density is given by a Dirac delta function\footnote{Our wave function is essentially the real part of the state discussed by G. Kaiser \cite{Kaiser:1990ww}. The author interprets his wave function as a coherent state, we show in this section however that this state is not a coherent state, it does not saturate the uncertainty relations. A similar state also appears in \cite{Wiese} but the authors do not normalize their states in a relativistically invariant manner which allows them to use the localization concept of non-relativistic quantum mechanics. However, by using wave functions that are not normalized in a covariant manner, they completely ignore the issue of relativistic position operators. Their results are nevertheless qualitatively similar to ours since our bilinear operator is the straightforward analogue of the position operator in non-relativistic quantum mechanics.}. For clarity we discuss here the wave packet that is peaked around the origin, both in position and momentum space. We also choose the initial time of the state such that $t=t_0=0$. The wave packet of the state in momentum space is given by,
\beq
f_+(\vec{k})= f_-(\vec{k}) = 4\pi \epsilon ~ e^{-\epsilon\|\vec{k}\|}.
\eeq
In position space the wave function is defined by,
\beq
\phi(x) =
(4\pi\epsilon) \int \frac{d^3k}{(2\pi)^3 \|\vec{k}\| } \left( e^{-i\|\vec{k}\|(t-i\epsilon)
+ i \vec{k}\cdot\vec{x}} +e^{i\|\vec{k}\|(t+i\epsilon)+ i \vec{k}\cdot\vec{x}}
\right),
\eeq
which gives, with $r=\|\vec{x}\|$,
\beq
\phi(x) = \frac{1}{\pi} \left(\frac{\epsilon}{-(t-i\epsilon)^2 - r^2} 
+ \frac{\epsilon}{-(t+i\epsilon)^2 - r^2}\right).
\eeq
The Hilbert Transform of the wave function is,
\beq
\phi_H(x) = \frac{1}{\pi} \left(\frac{i \epsilon}{-(t-i\epsilon)^2 - r^2} - \frac{i\epsilon}{-(t+i\epsilon)^2 - r^2}\right).
\eeq
At $t=0$ the probability density as defined by \eqref{Hilbertprobability} is,
\beq \label{t0prob}
P(x) = \half\phi(x) \on{\leftrightarrow}{\partial}_0 \phi_H(x) = \frac{4}{\pi} \frac{\epsilon^3}{[\epsilon^2+r^2]^3}.
\eeq
As observed in \cite{causalalternative} the probability density that corresponds to our state is perfectly localized in the limit where $\epsilon \rightarrow 0$,
\beq \label{problimit}
\lim_{\epsilon \rightarrow 0} P(x) = \delta^{(3)}(\vec{x}).
\eeq
The point we wish to make here is that this sense of localization corresponds precisely to the concept of localization as described by our bilinear Hermitian position operators. It is easy to see that since our wave packet in momentum space is real and our position operator is antisymmetric, that the average position of the wave function is zero. The width of the wave function in position space can be computed by our real and symmetric position squared operator.
In the $z$ direction we have that\footnote{Since the state that we discuss is spherically symmetric any other direction would yield the same result.}, 
\beq\label{zsqaverage}
\langle \hat{z}^2 \rangle = 
(4\pi \epsilon)^2\half \int \frac{\mathrm{d}^3k}{(2\pi)^3 \|\vec{k}\|} 
e^{-\epsilon\|\vec{k}\|}\left( (\on{\leftarrow}{\partial}_{k_z})^2 
+ (\on{\rightarrow}{\partial}_{k_z})^2\right) e^{-\epsilon\|\vec{k}\|} 
= (4\pi \epsilon)^2 \int \frac{\mathrm{d}^3k}{(2\pi)^3 \|\vec{k}\|} 
e^{-\epsilon\|\vec{k}\|} \partial^2_{k_z} e^{-\epsilon\|\vec{k}\|}. 
\eeq
This integral can be computed for arbitrary positive epsilon by introducing spherical coordinates in the momentum space variables, 
giving $\langle z^2 \rangle = \epsilon^2$. Since $\langle z \rangle = 0$ we have that the width of the wavepacket as defined by its standard deviation is simply 
given by,
\beq
\Delta z = \epsilon.
\eeq
From this it is clear that the wave packet becomes more and more localized as $\epsilon$ approaches zero. So indeed our symmetric position squared operator is perfectly compatible with the localization as described by \eqref{problimit}. This should come as no surprise since our position operators in momentum space simply correspond to the averages of the coordinates in the position space expressions. The expectation value $\langle \hat{z}^2 \rangle$ as computed in momentum space \eqref{zsqaverage} is simply equivalent  to the position space integral,
 \beq
 \langle \hat{z}^2 \rangle = \int d^3x~z^2~P(x) = \epsilon^2,
 \eeq
where $P(x)$ is the probability density given by \eqref{t0prob} and the integral is solved by going to spherical coordinates. We reiterate that the Newton-Wigner localized state does not posses a probability density since it is not normalizable and therefore does not correspond to a perfectly localized wavepacket even if we were to use our bilinear position operators. 
 
The momentum localization properties are described by the standard deviation of the momentum operators. Since our probability density in momentum space is even in the momenta the average momentum is zero. The momentum squared average can be computed and is given by,
\beq\label{zsqmomaverage}
\langle k_z^2 \rangle 
= (4\pi \epsilon)^2 \int \frac{\mathrm{d}^3k}{(2\pi)^3 \|\vec{k}\|} k^2_z e^{-2\epsilon\|\vec{k}\|} = \frac{1}{2 \epsilon^2}.
\eeq
The width of the wave packet in momentum space as given by the standard deviation is therefore,
\beq
\Delta k_z = \frac{1}{\sqrt{2}} \frac{1}{\epsilon}.
\eeq
The  phase space localization properties of the wave function are therefore independent of $\epsilon$,
\beq
\Delta z \Delta k_z = \frac{1}{\sqrt{2}},
\eeq
which exceeds the bound set by Heisenberg's uncertainty relation by a factor $\sqrt{2}$.
The state is therefore reasonably localized in phase space but is not optimally localized for any value of $\epsilon$. It is nonetheless perfectly localized in position space if $\epsilon$ tends to zero and becomes perfectly localized in momentum space in the limit where $\epsilon$ tends to infinity.

For $t\neq0$ the probability density is given by,
\beq \label{tfiniteprob}
P(x) 
= \frac{4}{\pi} \frac{\epsilon^3 (t^2+r^2 + \epsilon^2)}{[-(t-i\epsilon)^2 + r^2]^2[-(t+i\epsilon)^2 + r^2]^2}.
\eeq
The state can of course not be perfectly localized around $\vec{x}=0$ for arbitrary times by taking a limit. Interestingly however the state spreads but remains localized in the sense that is confined to the lightcone,
\beq \label{tfiniteeps0prob}
\lim_{\epsilon \rightarrow 0} P(x) 
= \frac{1}{4\pi r^2}[\delta(t-r)+\delta(t+r)].
\eeq
Observe that the magnitude of the probability decreases with the area of the spherical surface around the point of initial localization. Notice furthermore that the probability density is an expanding spherical wave for $t>0$ and a contracting spherical wave for $t<0$. This indicates that both advanced and retarded solutions to the field equation contribute to the probability density since the retarded solution of the field equation is an expanding spherical wave valid for $t>0$ and the advanced solution is a contracting spherical wave solution for $t<0$. This also fits perfectly with the half-advanced half-retarded nature of the Wheeler propagator. 

Note that in particular in classical electrodynamics one often only works with the retarded contribution. Without further input however such solutions violate conservation of energy rather dramatically, and in our case conservation of probability would be violated as well. Our probability density is however perfectly conserved for arbitrary $\epsilon$ and its limit yields the half-advanced half-retarded result  \eqref{tfiniteeps0prob}. 

So far we have only discussed the probability density, it is also illuminating to study the energy density. 
The total energy of our state is most easily calculated in momentum space where it has the simple form of an expectation value \eqref{energyexpectationvalue2}, for $t=0$ we have,
\beq \label{energyforumula}
E = 
(4\pi \epsilon)^2 \int \frac{\mathrm{d}^3k}{(2\pi)^3} e^{-2\epsilon\|\vec{k}\|} 
= \frac{1}{\epsilon}.
\eeq
This value is also valid for arbitrary times by energy conservation. From \eqref{energyforumula} we see that a perfectly localized particle seems possible in principle but in practice it would require an infinite amount of energy to do so. Again this agrees well with physical intuition based on Heisenberg's uncertainty principle. 

Note that in many respects the wave function that we discuss resembles a Gaussian. It is however difficult to compute the relativistic dynamics of a Gaussian wave packet while the time evolution of the wave function discussed here can be fully treated by analytical methods. The case of massive fields can also be treated in some detail, we chose the massless example however to illustrate our main points related to localization as clearly as possible.
 
%%%%%%%%%%%%%%%%%%%%%%%%%%%%%%%%%%%%%%%%%%%%%%%%%%%%%%%%%%%%%%%%%%%%
\section{Conclusions}
%%%%%%%%%%%%%%%%%%%%%%%%%%%%%%%%%%%%%%%%%%%%%%%%%%%%%%%%%%%%%%%%%%%%

We gave an interpretation of free quantum field theory as relativistic quantum mechanics by including negative frequency modes in the wave functions. The appropriate single particle Hilbert space for neutral scalar particles can be constructed from the probability density,
\beq \label{probabilityconlusion}
P(x)=\half\phi(x)  \on{\leftrightarrow}{\partial}_{0} \phi_H(x),
\eeq
where $\phi_H(x)$ denotes the Hilbert transform of $\phi(x)$ with respect to time, where we emphasize that $\phi_H(x)$ is real.
The probability density \eqref{probabilityconlusion} gives a wave function interpretation to the real Klein Gordon field so the Klein Gordon equation becomes the relativistic analogue of the Schr\"{o}dinger equation.  The propagator that replaces the Feynman propagator is the Wheeler propagator which is the half-retarded half-advanced propagator,
\beq
\phi(x) =  
\half\theta(t-t') \int \mathrm{d}^3 x' \Delta_W(x-x')   \on{\leftrightarrow}{\partial}_{t'} \phi(x')
+ \half\theta(t'-t)\int \mathrm{d}^3 x'  \phi(x'') \on{\leftrightarrow}{\partial}_{t'} \Delta_W(x''-x).
\eeq
Since the Wheeler propagator vanishes outside the light cone the dynamics of the real wave function $\phi(x)$ is manifestly causal. Moreover, since both our 
propagation law and our probability density contain real variables only implies that we are able to formulate quantum mechanics in position space using real 
variables only. Furthermore, we discussed the bilinear position operators,
\beq
\hat{x}^\mu \rightarrow  \frac{i}{2}\on{\leftrightarrow}{\eth}_{k_\mu}, \qquad 
\widehat{x^\mu x^\nu} \rightarrow
-\half\left(\on{\leftarrow}{\eth}_{k_\mu}\on{\leftarrow}{\eth}_{k_\nu}+\on{\rightarrow}{\eth}_{k_\mu} \on{\rightarrow}{\eth}_{k_\nu}\right).
\eeq
These operators are covariant, Hermitian, local and well defined for particles with spin such as photons. Therefore they are a straightforward extension of the non-relativistic position operators in the Schr\"{o}dinger picture, unlike operators related to the Newton-Wigner position operator. Interestingly, the bilinear nature of our operators makes the reality of their expectation values manifest. The antisymmetric operators imply that an imaginary part is taken and the symmetric operators imply that a real part is taken. Additionally, the bilinear space-time position operator combines quantum mechanics and special relativity in the sense that it allows the covariant Weyl-Heisenberg-Poincar\'e algebra to be realized both in momentum and position space. 

Particularly, our space-time position operator explicitly contains a time operator that can be exponentiated to a unitary operator. A standard argument due to Pauli that the time operator cannot be unitarily exponentiated is rooted in the fact that the energy is bounded from below. This problem is evaded simply because our time operator is dual to the frequency and not the energy. Furthermore, since our wave functions also include negative frequencies no problems with unitarity occur when our time operator is exponentiated.

Moreover, we have presented an explicit real solution to the Klein Gordon equation $\phi_\epsilon$ that can be perfectly localized in the sense that it has a limit where the width of its probability density becomes zero and the density itself becomes a Dirac delta function,
\beq
\lim_{\epsilon \rightarrow 0} \half\phi_\epsilon(x)  \on{\leftrightarrow}{\partial}_{0} \phi_\epsilon^H(x) = \delta^{(3)}(\vec{x}).
\eeq
We have also shown that the limit cannot be attained in practice however since the total energy contained in the state diverges, which is what one expects from physical intuition based on Heisenberg's uncertainty relations,
\beq
E_\epsilon = \int d^3x T_{00}(x)= \frac{1}{\epsilon}.
\eeq
The width of the probability density in position space is given by $\epsilon$,  which can be verified using our Hermitian bilinear position operators in momentum 
space. Hence, the degree of localization of a wave function can be obtained with our bilinear position operators and perfectly localized relativistic wave functions can be obtained in the limit of infinite energy. This should be contrasted with the ``localized state'' of Newton and Wigner \cite{Newton:1949cq} which is not 
normalizable and therefore does not describe an arbitrarily localized probability density.

Together with \cite{causalalternative} we offer a consistent first quantization of free quantum field theory because the covariant dynamics of the wave functions is causal and our position operator and probability density yield a local probabilistic interpretation. We expect a similar picture to hold also in perturbative interacting quantum field theory. Pair creation and other types of interactions will then play a role but those effects should be treated as small corrections to our localization scheme provided the coupling constant is small.  Again we contrast this with the Newton-Wigner localization scheme were particles are not even localizable in free quantum field theory. A consistent local interpretation seems especially useful in modern applications of quantum field theory where localization is important such as cavity quantum electrodynamics.

%ww soften argument + opsomming van maken
We are currently investigating the quantitative influence of interactions on the here described concept of localization. It would also be interesting to generalize our formalism to curved space-times.

\appendix
%%%%%%%%%%%%%%%%%%%%%%%%%%%%%%%%%%%%%%%%%%%%%%%%%%%%%%%%%%%%%%%%%%%%
\section{Conventional field theory as relativistic quantum mechanics} 
\label{Conventional}
%%%%%%%%%%%%%%%%%%%%%%%%%%%%%%%%%%%%%%%%%%%%%%%%%%%%%%%%%%%%%%%%%%%%
The relativistic analogue of the non-relativistic Schr\"{o}dinger wavefunction is the real Klein Gordon field since both describe particles without spin and electric charge. The real Klein Gordon field contains both negative and positive energy modes however and satisfies the second order Klein Gordon equation which make an interpretation of the real field as a wave function difficult.  The positive and negative frequency solutions separately satisfy first order equations however,
\beq
i \hbar \partial_t \phi_\pm(x)   
= \pm m c^2\sqrt{1-\left(\frac{\hbar}{m c}\right)^2\vec{\partial}^2}~\phi_\pm(x),
\eeq
which have the form of a Schr\"{o}dinger equation. This is especially clear from the non-relativistic approximation where the free particle Schr\"{o}dinger equation is recovered exactly, with the appropriate relativistic corrections, 
\beq
\pm i \hbar \partial_0 \:\phi_\pm(x) 
= 
m c^2 \phi_\pm - \frac{\hbar^2}{2 m} \vec{\partial}^2 \phi_\pm 
- \frac{1}{8} \frac{\hbar^4}{m^3 c^2} \vec{\partial}^4\phi_\pm-...\
\eeq
Since the positive frequency components of real fields are complex valued one can interpret these as wave functions by using a relativistic analogue of the ``wave function squared rule'' of Max Born\footnote{As acknowledged by Born himself in his Nobel lecture \cite{nobel}, Einstein was actually the first to attribute a statistical interpretation to the amplitude squared in the context of photons. It is amusing however that to this day the probabilistic interpretation of photons is far from clear, see for instance \cite{BialynickiBirula:1997am}.} in non-relativistic quantum mechanics \cite{MaxBorn, nobel}. 
\beq %\label{KGprobability}
P(x) = \phi_+^*(x) i \on{\leftrightarrow}{\partial}_{0} \phi_+(x). \nonumber
\eeq
whereas the Schr\"{o}dinger form of the Born rule is simply,
$
P_{Sch}(x) = \psi^*(x) \psi(x).
$ 
The Schr\"{o}dinger probability density, and therefore also the wave function, is not Lorentz covariant which is clear from the normalization condition,
\beq
\int d^3 x \psi^*(x) \psi(x) = 1.
\eeq
The integration element is not a Lorentz scalar but a four vector valued three form\footnote{The integration element is in fact $
d^3x^\mu = \frac{1}{4!} dx^\kappa \! \wedge \!dx^\lambda\! \wedge \!dx^\sigma \tilde{\varepsilon}_{\kappa \lambda \sigma }^{~~~~\mu} 
$
which in standard choice of coordinates, adapted to a spatial hyper-surface, reduces to $d^3x = dx^1dx^2dx^3$, see for instance the discussion on p. 134 of  Ryder \cite{Ryder:1985wq}.}, therefore the Schr\"{o}dinger probability density is not a Lorentz scalar, in fact it should be a vector valued current. The covariant nature of the probability density can be made manifest if normalize wave functions by a boundary integral,
\beq \label{KleinGordonnorm}
\int d^4x \partial^\mu (\phi_+^*(x) i \on{\leftrightarrow}{\partial}_{\mu} \phi_+(x))
=
 \int d^3x^\mu\phi_+^*(x) i \on{\leftrightarrow}{\partial}_{\mu} \phi_+(x) = 1,
\eeq
which with a standard choice of hypersurface and coordinates this leads to the probability density \eqref{KGprobability}. 
Note that $\phi_+(x)$, contrary to $\psi(x)$, is a Lorentz scalar. The relativistically invariant momentum space wave functions are obtained by the standard on shell Fourier transform,
\beq
\phi_+(x) 
= 
\int \frac{\mathrm{d}^4k}{(2 \pi)^3}\delta(k^2+m^2)\theta(k^0)f(k)e^{i k x},
\eeq
such that similar to the wave functions $\phi_+(x)$ the wave packets $f(k)$ are Lorentz scalars. The normalization condition on the wave packets $f(k)$  is simply the Fourier transform of \eqref{KleinGordonnorm} in standard coordinates,
\beq \label{KGnormalization}
\int \frac{d^4k}{2 \pi^3}\delta(k^2+m^2)\theta(k_0) f^*(k)  f(k)=
\int  \frac{\mathrm{d}^3k}{(2 \pi)^3 2\omega_k} f^*_+(\omega_k,\vec{k})  f_+(\omega_k,\vec{k}) = 1,
\eeq
where $\omega_k = \sqrt{\vec{k}^2+m^2}$ and $f_+(\omega_k,\vec{k})$ is a Lorentz invariant positive frequency wavepacket. 
The positive frequency contribution $f_+(\omega_k,k)$ is a wave function in the sense that it gives a probability density by the following Born rule,
\beq
P(\omega_k,\vec{k},t) 
= 
\frac{1}{(2 \pi)^3 2\omega_k}  f^*_+(\omega_k,\vec{k},t)  f_+(\omega_k,\vec{k},t).
\eeq
For a text book discussion on wave packets in quantum field theory consult for example section 3.1. of \cite{DeWit:1986it}, section 3.4.  of \cite{Brown:1992db} or section 4.5. of \cite{Peskin:1995ev}. To summarize one might say that in the quantum mechanical interpretation of conventional quantum field theory \emph{quantization is essentially the extraction of the positive frequency components from the fields and enforcing the normalization condition \eqref{KGnormalization} as required by a probabilistic interpretation}. Note that this can be done without promoting the fields to operators. So from this point of view the operator language merely is a convenient tool to organize multi-particle  wave functions but \emph{not} essential for quantization. We used the term quantization in the restricted sense of providing the space of solutions to the equations of motion with an appropriate Hilbert space structure, see \eqref{KleinGordonnorm} and \eqref{KGnormalization}, associated to the positive frequency part of the fields.

We stress however the well known fact that the above discussed quantum mechanical picture of standard quantum field theory is not physical however as issues with causality and non-locality are present. Because of these problems the first quantized interpretation is most often abandoned. One should realize that nonetheless it is the interpretation used in textbook methods to convert scattering amplitudes to scattering probabilities and cross sections. The modern interpretation of quantum field theory is not the one that is used in scattering problems. Most often quantum field theory is given a functional interpretation, the field is not considered to be a wave function but an infinite set of oscillator degrees of freedom. It is the purpose of this paper to show that the field \emph{can} be interpreted as a relativistic wave function provided one includes its negative frequency modes and uses a suitable position operator.  
%
%%%%%%%%%%%%%%%%%%%%%%%%%%%%%%%%%%%%%%%%%%%%%%%%%%%%%%%%%%%%%%%%%%%%
\subsubsection{Time evolution violates Einstein causality} \label{noncausality}
%%%%%%%%%%%%%%%%%%%%%%%%%%%%%%%%%%%%%%%%%%%%%%%%%%%%%%%%%%%%%%%%%%%%
%
One of the problems with interpreting quantum field theory as a first quantized theory is that the time evolution of positive frequency solutions to the Klein Gordon equation, which are meant to be interpreted as wave functions,  do not respect Einstein causality \cite{causalalternative, Hegerfeldt:1998ar, Hegerfeldt:1974qu}. This can be seen by realizing that the Feynman propagator from standard perturbation theory governs the time ordered evolution of the positive energy modes,
\begin{subequations}
\label{Feynmanpropagationlaw}
\begin{eqnarray}
\theta(t-t')\phi_+(x) 
&=& i \int \mathrm{d}^3 x'  \Delta_F(x-x')   \on{\leftrightarrow}{\partial}_{t'} \phi_+(x'), \label{Feynmanpropagatoraa}\\
\theta(t'-t)\phi_+^*(x) 
&=& i \int \mathrm{d}^3 x' \phi_+^*(x') \on{\leftrightarrow}{\partial}_{t'} \Delta_F(x-x')\, \label{Feynmanpropagatorbb}.
\end{eqnarray}
\end{subequations}
Formula \eqref{Feynmanpropagatoraa} shows that the Feynman propagator is the retarded propagator for positive frequency fields and \eqref{Feynmanpropagatorbb} shows that it is the advanced propagator for the complex conjugate positive frequency field. Notoriously, the Feynman propagators possesses contributions outside of the light-cone which directly implies non-causal spreading of a wave function $\phi_+$ by the propagation law \eqref{Feynmanpropagationlaw}, for an explicit example see \cite{causalalternative}. The contributions that violate causality are clear from the explicit form of the Feynman propagator in position space,
\beq \label{Feynmanpositionspace}
\Delta_F(x) =
\theta(-x^2)\frac{m}{8 \pi \sqrt{-(x^2 +i \varepsilon)}}H^{(2)}_1\left(m\sqrt{-(x^2+i\varepsilon)}\right)+ 
i\theta(x^2)\frac{m}{4\pi^2 \sqrt{x^2 +i \varepsilon}}K_1\left(m\sqrt{x^2+i\varepsilon}\right),
\eeq
where the second term is nonzero outside of the lightcone, which follows from the theta function $\theta(x^2) = \theta(-c^2t^2 + \vec{x}^2)$ which enforces spacelike contributions only. 
%

%%%%%%%%%%%%%%%%%%%%%%%%%%%%%%%%%%%%%%%%%%%%%%%%%%%%%%%%%%%%%%%%%%%%
\section{Causality from negative energies} \label{causality}
%%%%%%%%%%%%%%%%%%%%%%%%%%%%%%%%%%%%%%%%%%%%%%%%%%%%%%%%%%%%%%%%%%%%

Perhaps less often mentioned then is deserved but emphasized in our recent work \cite{causalalternative} is the fact that the \emph{real part of the Feynman propagator is perfectly causal} and vanishes outside of the Lightcone. We call this propagator the Wheeler propagator following \cite{Bollini:1998hj} due to its implicit appearance in the Wheeler Feynman absorber interpretation of classical electromagnetism \cite{Wheeler:1945ps}. A more modern application of the Wheeler propagator is the perturbative description of the classical gravitational problem of pointlike particles \cite{Damour:2001bu,Damour:2008ji}. It is also known as the principal part propagator \cite{Greiner:1996zu} since it's momentum space definition involves a principal part \cite{Greiner:1996zu, Birrell:1982ix,Itzykson:1980rh}. The operational significance of the Wheeler propagator as stressed in  \cite{causalalternative} is that it governs the time ordered time evolution of the whole real field and not just its positive frequency part, the Wheeler propagator applied to the field gives
\begin{eqnarray}{}\label{Wheelerpropagator}
\half \epsilon(t-t')\phi(x) &=&  \int \mathrm{d}^3 x'  \Delta_W(x-x')   \on{\leftrightarrow}{\partial}_{t'} \phi(x').
\end{eqnarray}
While the Feynman propagator is related to the a-causal positive frequency Wightman function, the Wheeler propagator is related to the causal commutator function\footnote{The commutator two-point function is also referred to as the spectral propagator \cite{Berges:2004yj} in the  Schwinger Keldysch, closed time path or in-in formalism popular in cosmological applications \cite{Calzetta:1986ey,Jordan:1986ug, Chou:1984es, Weinberg:2005vy}.}  by time ordering, and is therefore also related to the ``classical'' advanced and retarded propagators,
\beq \label{Wheelercommutator}
\Delta_W(x-x') = 
\half\theta(t-t')\Delta_C(x-x')+ \half\theta(t'-t)\Delta_C(x'-x), \qquad \Delta_W 
= \half\Delta_R + \half\Delta_A,
\eeq
where the commutator can be given the integral representation,
\beq \label{commutatorintegral}
\Delta_C(x-x') = 
\half\int \frac{\mathrm{d}^4k}{(2 \pi)^3} \delta(k^2+m^2)\epsilon(k_0)e^{ik(x-x')}\,,
\eeq
and $\epsilon(k_0)$ is the sign function which written in terms of Heaviside functions is $\theta(k_0) - \theta(-k_0)$.The integral representation of the Wheeler propagator is,
\beq \label{principalpartpropagator}
P\Big[\int \frac{\mathrm{d}^4k}{(2 \pi)^4} \frac{1}{k^2 + m^2 }e^{ik(x-x')}\Big],
\eeq
where $P$ denotes the principal value.The explicit form of the Wheeler propagator which follows from \eqref{Wheelercommutator} and \eqref{commutatorintegral} is,
\beq \label{Wheelerpositionspace}
\Delta_W(x)  =
\frac{1}{2\pi}\delta(-x^2) 
+ \theta(-x^2)\frac{m}{4 \pi \sqrt{-x^2}}J_1\left(m\sqrt{-x^2}\right).
\eeq
From this expression it is manifestly clear that the Wheeler function cannot  propagate fields outside the light-cone. Note that \eqref{Wheelerpositionspace} is the real part of the regulated expression of the Feynman propagator \eqref{Feynmanpositionspace}, and  \eqref{principalpartpropagator} is the real part of the standard integral representation for the Feynman propagator, in both cases the regulator $\varepsilon$ is taken to zero.

Also note that the Wheeler propagator, not the Feynman propagator, is the natural Green's function of a real scalar field. This is rooted in the fact that Wheeler propagator is a \emph{real} and symmetric solution to the inhomogeneous Klein Gordon equation.

%%%%%%%%%%%%%%%%%%%%%%%%%%%%%%%%%%%%%%%%%%%%%%%%%%%%%%%%%%%%%%%%%%%%
\subsection{The in-out boundary conditions implied by the Wheeler propagator}
%%%%%%%%%%%%%%%%%%%%%%%%%%%%%%%%%%%%%%%%%%%%%%%%%%%%%%%%%%%%%%%%%%%%

A perhaps more intuitive understanding of the half-advanced, half-retarded nature of the Wheeler propagator can be gained from the following form of the propagation law,
\beq 
\phi(x) =  \half\theta(t-t')\int \mathrm{d}^3 x' \Delta_W(x-x')   \on{\leftrightarrow}{\partial}_{t'} \phi(x')
+ \half\theta(t'-t)\int \mathrm{d}^3 x'  \phi(x'') \on{\leftrightarrow}{\partial}_{t'} \Delta_W(x''-x),
\eeq
which is equivalent to \eqref{Wheelerpropagator}. The physical interpretation of this equation is as follows, half of the field $\phi(x)$ at time $t$ is specified by initial conditions and the other half of the field is specified by ``final conditions''. The initial condition is the specification of a real field $\phi(x')$ at a time $t'$ earlier than $t$ and the final condition is the specification of a field $\phi(x'')$ at a time $t'$ later than $t$. The use of initial and final conditions on the field implies the use of two Dirichlet boundary conditions as opposed to the half-Dirichlet, half-Neumann Cauchy boundary conditions which are standard when viewing the Klein Gordon field as a ``classical'' field. In the context of quantum cosmology the Dirichlet boundary conditions are referred to as ``in-out'' boundary conditions and the Cauchy boundary conditions are referred to as ``in-in'' boundary conditions, see for instance \cite{Weinberg:2005vy}.

The Feynman propagator also corresponds to a choice of ``in-out'' boundary conditions. Unlike our ``in-out'' boundary conditions which are compatible with the Wheeler propagator,  the Feynman Stueckelberg boundary conditions are complex even for real fields. The initial conditions in the standard Feynman Stueckelberg formalism are not specified by the real field $\phi(x) = \phi_+(x)+\phi_-(x)$ but by the positive frequency solutions $\phi_+(x)$ only and the final conditions are given by $\phi^*_+(x)$. Equivalently, by the identity for real fields $\phi_-(x) = \phi^*_+(x)$ one can add the negative frequency solutions $\phi_-(x)$ to the \emph{final} conditions and their complex conjugates  $\phi^*_-(x)$  to the initial conditions. This corresponds to Feynman's frequently misinterpreted statement that negative frequency solutions ``travel back time''. Note that with our boundary conditions  there is no difference in the ``direction of propagation'' between positive and negative frequency fields. They only appear in their real combination and therefore both are given half-advanced, half-retarded boundary conditions.

%%%%%%%%%%%%%%%%%%%%%%%%%%%%%%%%%%%%%%%%%%%%%%%%%%%%%%%%%%%%%%%%%%%%
\subsection{Negative frequency modes have positive energy}
%%%%%%%%%%%%%%%%%%%%%%%%%%%%%%%%%%%%%%%%%%%%%%%%%%%%%%%%%%%%%%%%%%%%

Historically, 
%it was argued that all sorts of stability problems arise if one would include the negative frequency modes. 
negative frequency modes were thought to lead  stability problems
 due to the misconception that negative frequency modes have negative energy. The lore was that negative frequency wave functions would not allow a true ground state of lowest energy so that the vacuum would  decay indefinitely to lower energy states. Dirac's sea of occupied negative energy states is a famous attempt to resolve this problem for Fermions. In the modern Feynman and Stueckelberg approach both Bosonic as Fermionic negative frequency states are rewritten as positive frequency states with advanced boundary conditions. 

We do not agree with the negative sentiments towards negative frequency modes. As stressed in \cite{causalalternative} the essential point is that the energy contained in a field is \emph{not} given by the energy that appears in the dispersion relation, the true energy contained in a wave function is given by the energy momentum tensor,
\begin{align} \label{energyexpectationvalue}
E& = \int d^3x^0 T_{00}  =
\half \int d^3x (\partial_{0} \phi \partial_{0}\phi 
+ \vec{\partial}\phi \cdot \vec{\partial} \phi  + \mu^2  \phi^2),\\
 &=\half \int \frac{d^4k}{(2\pi)^3} \delta(k^2+m^2)~ |k_0| |f(k)|^2.  \label{energyexpectationvalue2}
\end{align}
Note the absolute value operation on $k_0$, it directly implies that the \emph{negative frequency modes do not have negative energy}. We therefore prefer to use the terminology ``negative frequency'' solutions as opposed to ``negative energy'' solutions\footnote{In the literature there do appear attempts to add modes whose energy \emph{is} considered to be negative \cite{Moffat:2005ip, Linde:1988ws, Kaplan:2005rr}. We do not follow such an approach however.}. Furthermore, notice that the on-shell formula for the energy \eqref{energyexpectationvalue2} has the form of a quantum mechanical expectation value. 

%%%%%%%%%%%%%%%%%%%%%%%%%%%%%%%%%%%%%%%%%%%%%%%%%%%%%%%%%%%%%%%%%%%%
\subsection{Negative frequency modes and the probability density} \label{negfreqprob}
%%%%%%%%%%%%%%%%%%%%%%%%%%%%%%%%%%%%%%%%%%%%%%%%%%%%%%%%%%%%%%%%%%%%

In this section we summarize our discussion in \cite{causalalternative} which was inspired by Halliwell and Ortiz \cite{Halliwell:1992nj} and define a probability density on the space of real fields. As written in \eqref{energyexpectationvalue} the energy of a scalar field can be interpreted as a quantum expectation value. Hence, the energy momentum tensor suggests the following normalization condition for the single particle wave function\footnote{This form of the innerproduct is discussed by Halliwell and Ortiz in \cite{Halliwell:1992nj} who were inspired by work of Henneaux and Teitelboim  \cite{Henneaux:1982ma}. The innerproduct also appears in a series of papers by P. Morgan where a classical stochastic interpretation is given, see e.g. \cite{Morgan:2009wa}.},
\beq \label{covariantnormmomentum}
\half\int \frac{d^4k}{2 \pi^3}\delta(k^2+m^2) |f(k)|^2 = 1,
\eeq
which is almost the same as the standard normalization of the Klein Gordon field \eqref{KGnormalization} except that no Heaviside function is employed to restrict the integral to positive frequencies. Consequently, a Lorentz invariant version of Born's rule for the probability density follows that includes negative frequency wave modes,
\beq \label{momentumprobability}
P(k) = 
\half\int d k_0 \delta(k^2+m^2) |f(k)|^2 
=\half \frac{1}{(2 \pi)^3 2\omega_k} \left( |f_+(\vec{k})|^2 +  |f_-(\vec{k})|^2\right).
\eeq
In position space this gives a negative frequency completion of the ``Klein Gordon Born-rule'' \eqref{KGprobability} that despite the presence of negative frequencies is positive definite,
\beq %\label{plusminprobability}
P(x) =
\half\left( \phi_+^*(x) i \on{\leftrightarrow}{\partial}_{0} \phi_+(x) 
- \phi_-^*(x)  i \on{\leftrightarrow}{\partial}_{0}  \phi_-(x)\right).\nonumber
\eeq
This formula is merely a rewritten form of the Klein Gordon form of the probability density since $\phi_- = \phi^*_+$ for real fields. Hence, adding negative frequencies does not change the probabilities of standard quantum field theory. Inspired by \cite{Halliwell:1992nj} we can write the probability density equivalently as,
\beq %\label{Hadamardprobabilitydensity}
P(x) = 
\half\int d^3x' 
\phi(x) \on{\leftrightarrow}{\partial }_{t} \Delta_H(x,x') \on{\leftrightarrow}{\partial }_{t'} \phi(x')\Big|_{t=t'}, \nonumber
\eeq
where the Hadamard function, sometimes referred to as statistical ``propagator'' and related to the anti-commutator, is given by,
\beq
\Delta_H(x-x')  = -\int \frac{\mathrm{d}^4k}{(2 \pi)^3} \delta(k^2+m^2) e^{ik(x-x')}\,,
\eeq
where we introduced an overall minus sign to avoid a minus sign in the definition of the probability density \eqref{Hadamardprobabilitydensity}. Note that the Hadamard form of the probability density \eqref{Hadamardprobabilitydensity} is equivalent to the Klein Gordon density \eqref{KGprobability}. The essential difference between the two representations is however that in the Hadamard probability the real fields themselves are relevant and not the positive or negative frequency contributions separately. This shows that the causal Wheeler propagator can be used instead of the a-causal Feynman propagator to compute the dynamics of wave functions and probabilities. Notice that the Hadamard propagator and its time derivative only appear with equal time arguments, therefore it does not influence the causal propagation of the probability density. The Hadamard form of the probability density can directly be compared to the Klein Gordon probability density in the following form,
\beq \label{KGinnerproductWightman}
P(x )= 
\int d^3x' 
\phi(x)\on{\leftrightarrow}{\partial}_0 \Delta^+(x-x') \on{\leftrightarrow}{\partial}_{0'} \phi(x')\Big|_{t=t'},
\eeq
This expression reduces to the form familiar from the position space probability density that we introduced at the beginning of this paper \eqref{KGprobability}. To establish the equivalence of \eqref{KGinnerproductWightman} and \eqref{KGprobability} one uses the fact that the positive frequency Wightman function is a propagator for positive frequency fields,
\beq \label{Wightmanpropagation}
\phi_+(x) = \int d^3x'\Delta^+(x-x') \on{\leftrightarrow}{\partial}_{0'} \phi_+(x'),
\eeq
We emphasize that in standard quantum field theory the Wightman, or including time ordering the Feynman, propagator plays a dual role,it both gives the time evolution of the wave function by means of its propagation law \eqref{Wightmanpropagation} \emph{and} it is used to define the innerproduct  \eqref{KGinnerproductWightman}. If one chooses however to work with real fields as wave functions, the time evolution and the probability density are determined by two different two-point functions. The Hadamard function defines the probability density \eqref{Hadamardprobabilitydensity} and the time evolution is governed by the commutator two point function, or  its time ordered counterpart the Wheeler propagator \eqref{Wheelerpropagator}. does not satisfy a law similar to \eqref{Wightmanpropagation}. The commutator two point function on the other hand \emph{is} the appropriate (non time ordered) propagator for real fields.
%%%%%%%%%%%%%%%%%%%%

%%%%%%%%%%%%%%%%%%%%%%%%%%%%%%%%%%%%%%%%%%%%%%%%%%%%%%%%%%%%%%%%%%%%
\section*{Acknowledgments}
%%%%%%%%%%%%%%%%%%%%%%%%%%%%%%%%%%%%%%%%%%%%%%%%%%%%%%%%%%%%%%%%%%%%
The author would like to thank J. Koksma for his support and  T. Prokopec  for a critical reading of the manuscript. W.W. would furthermore like to thank  S. Zohren, T. Jonsson, S. Stefansson for discussions at an earlier stage.

%%%%%%%%%%%%%%%%%%%%%%%%%%%%%%%%%%%%%%%%%%%%%%%%%%%%%%%%%%%%%%%%%%%%

%%%%%%%%%%%%%%%%%%%%%%%%%%%%%%%%%%%%%%%%%%%%%%%%%%%%%%%%%%%%%%%%%%%%

\end{document}